\documentclass[11pt]{article}

\usepackage[utf8]{inputenc}
\usepackage[T1]{fontenc}
\usepackage{amsmath,amssymb,amsthm}
\usepackage{graphicx}
\usepackage{hyperref}
\usepackage{booktabs}
\usepackage{cite}
\usepackage{authblk}
\usepackage[margin=2cm]{geometry}
\usepackage{tikz}
\usetikzlibrary{arrows.meta,positioning,shapes.geometric,calc,fit,backgrounds}
\usepackage{pgfplots}
\pgfplotsset{compat=1.18}
\usepackage{algorithm}
\usepackage{algpseudocode}
\usepackage{enumitem}
\usepackage{xcolor}
\usepackage{caption}
\usepackage{subcaption}
\usepackage{microtype}

\newtheorem{definition}{Definition}

\newtheorem{proposition}{Proposition}
\newtheorem{hypothesis}{Hypothesis}
\newtheorem{remark}{Remark}

\title{\textbf{Explainable PQC: A Layered Interpretive Framework\\for Post-Quantum Cryptographic Security Assumptions}}

\author[1]{Daisuke Ishii}
\author[2]{Rizwan Jahangir}
\affil[2]{NUST Business School, NUST, Islamabad, Pakistan}
\affil[1,2]{Kiara Inc., Tokyo, Japan}

\date{}

\begin{document}

\maketitle

\begin{abstract}
This paper studies how post-quantum cryptographic (PQC) security assumptions can be represented and communicated through a structured, layered framework that is useful for technical interpretation but does not replace formal cryptographic proofs.

We propose ``Explainable PQC,'' an interdisciplinary framework connecting three layers: (1)~a complexity-based interpretive model that distinguishes classical security, quantum security, and reduction-backed hardness, drawing on computational complexity classes as supporting language; (2)~an exploratory mathematical investigation applying combinatorial Hodge theory and polyhedral geometry to study structural aspects of lattice hardness; and (3)~an empirical experimentation platform, implemented in Julia, for measuring the behavior of lattice basis reduction algorithms (LLL, BKZ) in low-dimensional settings. The motivating case study throughout the paper is lattice-based PQC, including ML-KEM (FIPS 203) and ML-DSA (FIPS 204).

The contribution of this paper is conceptual and organizational: it defines a layered interpretive framework, clarifies its scope relative to formal cryptographic proofs and reduction-based security arguments, and identifies mathematical and implementation-level directions through which PQC security claims may be more transparently communicated. This paper does not claim new cryptographic hardness results, new attacks, or concrete security parameter estimates.
\end{abstract}


\vspace{1em}

\section{Introduction}

\subsection{Problem Statement}

The security of post-quantum cryptographic schemes rests on computational assumptions---such as the hardness of the Learning With Errors (LWE) problem~\cite{regev2009} and approximate variants of the Shortest Vector Problem (SVP)~\cite{peikert2016}---that are well-founded in complexity theory but opaque to non-specialist audiences. The NIST post-quantum standardization process has produced concrete standards---FIPS 203 (ML-KEM)~\cite{nist_fips203} and FIPS 204 (ML-DSA)~\cite{nist_fips204}, derived from CRYSTALS-Kyber and CRYSTALS-Dilithium respectively---yet the gap between the formal security arguments underlying these schemes and the understanding available to industry practitioners remains wide.

This paper addresses the following question: \emph{How can PQC security assumptions be communicated through a framework that aligns complexity-theoretic reasoning, mathematical structure, and implementation-level evidence, without claiming new hardness proofs?}

\subsection{Research Questions}

This work explores three specific questions:
\begin{description}[leftmargin=*]
    \item[RQ1.] Can cryptographic security assumptions be represented through a layered interpretive model that separates complexity-theoretic reasoning, mathematical structure, and implementation behavior?
    \item[RQ2.] Do geometric decompositions of lattice structures---specifically, fan decompositions and local generation theorems---suggest new perspectives for understanding algorithmic difficulty in lattice problems?
    \item[RQ3.] Can empirical experimentation in low-dimensional lattices illustrate scaling behavior relevant to understanding post-quantum cryptographic assumptions?
\end{description}

\subsection{Background: The Gap Between Quantum Threats and Trust}

The advancement of quantum computing technology, particularly the increasing plausibility of implementing Shor's algorithm~\cite{shor1997}, poses a significant threat to industries built upon current cryptographic infrastructure. RSA and elliptic-curve cryptography are vulnerable to quantum polynomial-time attacks on integer factorization and the discrete logarithm problem. In response, NIST has finalized its first set of PQC standards~\cite{nist_fips203,nist_fips204}, selecting lattice-based schemes whose security is grounded in standard lattice problems and reduction-based assumptions, including LWE and approximate lattice problems~\cite{regev2009,micciancio2009,peikert2016}.

However, within public and industrial discussions regarding the timeline of quantum advantage, there exists a mixture of excessive pessimism and optimism, leading to situations where rational decision-making is hampered. The primary source of this confusion is that the security foundations of PQC depend on highly sophisticated mathematical models, creating a structural separation between cryptographic researchers and industrial users.

\subsection{Contributions}

This paper makes the following contributions:
\begin{enumerate}[leftmargin=*]
    \item It defines a \textbf{layered interpretive framework} for PQC security communication and clearly delimits that framework from formal proof-based cryptographic security.
    \item It proposes a \textbf{complexity-oriented vocabulary} for distinguishing classical vulnerability, quantum vulnerability, and reduction-backed post-quantum assumptions.
    \item It identifies a concrete \textbf{mathematical program}, based on combinatorial Hodge theory and fan decompositions, for studying structural aspects of lattice hardness, and clearly separates established mathematical results from interpretive claims and open problems.
    \item It presents an \textbf{empirical experimentation platform} implemented in Julia that illustrates the rapid growth of exact lattice-solver computation cost in the tested low-dimensional regime. This layer is not intended as a security evaluation of deployed cryptographic systems, but as a controlled experimental environment for illustrating algorithmic scaling behavior in lattice reduction problems.
    \item It integrates these perspectives into a \textbf{single explanatory model} intended to connect cryptographic theory, mathematical structure, and implementation evidence for interpretability purposes.
\end{enumerate}

\subsection{Novelty}

The novelty of this paper is not a new hardness theorem or a new attack, but the formulation of a layered interpretive framework that places complexity-theoretic reasoning, exploratory lattice geometry, and implementation-level evidence into a single explanatory structure. To our knowledge, these three perspectives are rarely presented as a unified model for communicating PQC security assumptions. While each individual component draws on established literature, their integration into a coherent interpretive framework---with explicit scope limitations---constitutes the primary contribution.

\subsection{Related Work and Positioning}

This work intersects three established lines of research, but is distinct from each.

\paragraph{Formal lattice cryptography.} The theoretical foundations of lattice-based PQC are established through worst-case to average-case reductions. Regev~\cite{regev2009} proved that LWE is as hard as worst-case lattice problems (under quantum reductions), and Micciancio and Regev~\cite{micciancio2009} surveyed the landscape of lattice-based constructions. Peikert~\cite{peikert2016} provided a comprehensive treatment of SIS, LWE, and their ring-based variants. The NIST standardization process~\cite{nist_fips203,nist_fips204} produced concrete parameter sets and implementation specifications. Our work does not contribute new reductions, proofs, or parameter analysis; it takes these results as given and asks how they can be communicated more transparently.

\paragraph{Complexity theory and cryptography.} The relationship between computational complexity classes and cryptographic security is well studied. Aaronson~\cite{aaronson2010} established oracle results separating BQP from the polynomial hierarchy. The exact relationship between BQP and NP remains unknown, and the hardness assumptions used in deployed PQC concern approximate (not NP-hard exact) variants of lattice problems. Our complexity-based interpretive layer acknowledges these nuances explicitly, using complexity classes as a \emph{communication language} rather than as formal security criteria.

\paragraph{Interpretability and assurance communication.} While ``Explainable AI'' has become a major research area, there is no established ``Explainable PQC'' literature. Our work is closest in spirit to conceptual assurance frameworks and interdisciplinary security communication efforts that aim to bridge the gap between formal security arguments and practitioner understanding. Unlike formal verification or standards certification efforts, our objective is interpretive alignment rather than machine-checkable assurance. We propose a structured model for this communication rather than a formal verification system.

\begin{figure}[t]
\centering
\begin{tikzpicture}[
    layer/.style={draw, thick, rounded corners=4pt, minimum width=10cm, minimum height=1.6cm, fill=#1!8, align=center},
    note/.style={draw, dashed, rounded corners=3pt, fill=gray!5, text width=4.5cm, align=center, font=\small},
    arr/.style={-{Triangle[length=5pt]}, thick, gray},
]
\node[layer=blue] (L1) at (0,0) {\textbf{Layer 1: Complexity-Based Interpretation}\\{\small Classical vulnerability $\mid$ Quantum vulnerability $\mid$ Reduction-backed assumptions}};

\node[layer=orange] (L2) at (0,-2.5) {\textbf{Layer 2: Mathematical Structure}\\{\small Combinatorial Hodge theory $\mid$ Fan decompositions $\mid$ Lattice geometry}};

\node[layer=green!60!black] (L3) at (0,-5) {\textbf{Layer 3: Empirical Implementation}\\{\small Julia toolkit $\mid$ LLL / exact solver benchmarks $\mid$ Dimension-scaling experiments}};

\draw[arr] (L1.south) -- (L2.north) node[midway, right, font=\scriptsize, gray] {informs};
\draw[arr] (L2.south) -- (L3.north) node[midway, right, font=\scriptsize, gray] {motivates};

\node[note] (scope) at (7.5,-2.5) {\textbf{Scope Limitation}\\[2pt] Not a formal proof system.\\Intended for interpretability\\and communication.};
\draw[dashed, gray, thick] (L1.east) -- (scope.north west);
\draw[dashed, gray, thick] (L3.east) -- (scope.south west);

\end{tikzpicture}
\caption{Architecture of the Explainable PQC framework. Three layers connect complexity-theoretic interpretation, mathematical structure, and empirical evidence. The framework is explicitly scoped as an interpretive and communication tool, not a formal proof system.}
\label{fig:architecture}
\end{figure}
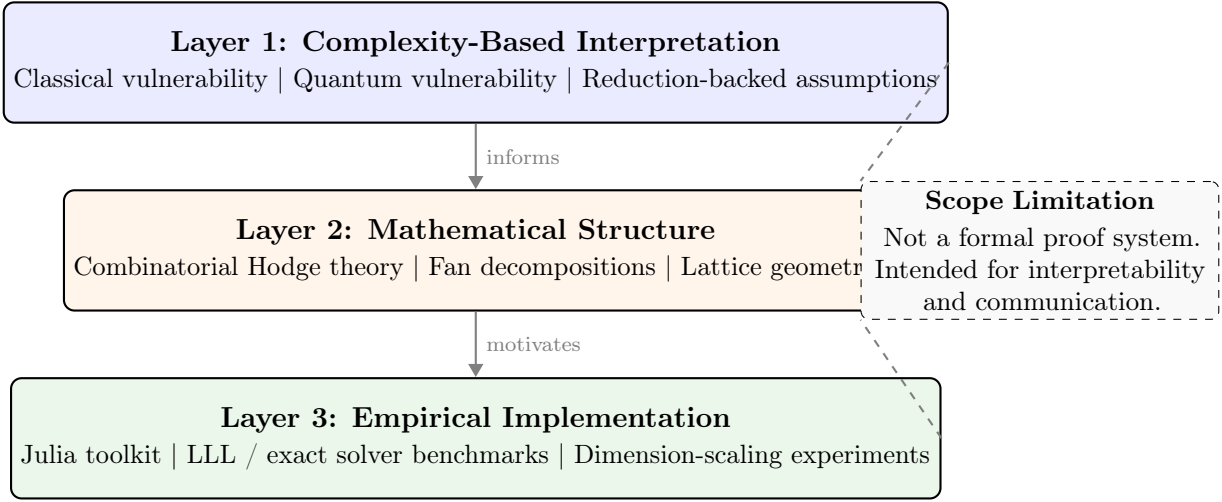

\subsection{Paper Organization}

Section~\ref{sec:theory} presents the three-part security interpretation (classical, quantum, and reduction-backed). Under this interpretive framework, conventional schemes such as RSA are characterized as classically secure but quantum-vulnerable, while lattice-based PQC schemes are conjectured to resist both classical and known quantum attacks, with their security grounded in standard lattice problems and reduction-based assumptions, including LWE and approximate lattice problems~\cite{regev2009}. We note that the exact relationship between BQP and NP remains open~\cite{aaronson2010}, and that deployed PQC relies on \emph{approximate} lattice problem hardness rather than NP-hard exact variants.

Section~\ref{sec:math} describes our exploratory mathematical investigation applying combinatorial Hodge theory~\cite{jahangir2025} and the local generation theorem~\cite{jahangir2026} to lattice geometry. Section~\ref{sec:implementation} details the empirical layer. Section~\ref{sec:limitations} discusses threats to validity. Section~\ref{sec:conclusion} summarizes contributions and outlines future directions.

\section{Three-Part Security Interpretation: Classical, Quantum, and Reduction-Backed}
\label{sec:theory}

We introduce the ``Classical--Quantum--Reduction Security Interpretation'' as a conceptual tool for interpreting the security posture of cryptographic schemes through the lens of computational complexity. This framework is not intended as a formal security proof, a substitute for cryptographic reduction proofs, or a concrete security parameter analysis. Rather, it serves as a structured communication model for a broader audience.

\begin{proposition}[Scope of the Interpretive Framework]
\label{prop:scope}
The Classical--Quantum--Reduction Security Interpretation is an interpretive classification tool. It does not by itself imply cryptographic security, average-case hardness, reduction-based guarantees, or concrete parameter security. Its purpose is to provide a structured vocabulary for discussing the \emph{type} of computational hardness that different cryptographic schemes rely upon.
\end{proposition}

\begin{definition}[Explainable PQC Framework]
\label{def:framework}
Let $\mathcal{C}$ denote the set of cryptographic constructions under consideration, and let $\mathcal{A}$ denote the set of computational assumptions on which their security relies. An Explainable PQC framework is a mapping
\[
\Phi : \mathcal{C} \rightarrow (S_c, S_q, S_r)
\]
where $S_c \in \{\text{Yes}, \text{No}, \text{Unknown}\}$ indicates resistance to known classical polynomial-time algorithms, $S_q \in \{\text{Yes}, \text{No}, \text{Unknown}\}$ indicates resistance to known quantum polynomial-time algorithms, and $S_r \in \{\text{Yes}, \text{No}\}$ indicates whether the construction is supported by reduction-based hardness arguments from established problems (e.g., worst-case to average-case reductions for LWE~\cite{regev2009}). The triple $(S_c, S_q, S_r)$ forms the \emph{interpretive security profile} of the cryptographic scheme.
\end{definition}

The purpose of $\Phi$ is interpretive rather than predictive: it summarizes how a scheme is presently positioned relative to known classical attacks, known quantum attacks, and reduction-based hardness arguments. It does not predict future attack developments or provide quantitative security margins.

\subsection{Interpretive Model}

The purpose of the framework is to provide a communication model with three outputs: (i)~whether a scheme's hardness assumption is broken by known classical algorithms, (ii)~whether it is broken by a known quantum polynomial-time algorithm, and (iii)~whether its security is backed by standard reduction-based arguments from established hard problems. Table~\ref{tab:safety_check} summarizes this classification.

\begin{table}[t]
\centering
\caption{Three-part security interpretation: Conceptual classification for cryptographic schemes.}
\label{tab:safety_check}
\begin{tabular}{@{}lp{5cm}p{5.5cm}@{}}
\toprule
\textbf{Indicator} & \textbf{Criterion} & \textbf{Interpretation} \\
\midrule
Classically Secure & No known polynomial-time classical algorithm & If No: potentially vulnerable to classical attacks \\
\addlinespace
Quantum Secure & No known polynomial-time quantum algorithm & If No: potentially vulnerable to quantum attacks (e.g., Shor's algorithm~\cite{shor1997}) \\
\addlinespace
Reduction-Backed & Security supported by worst-case to average-case reductions from standard hard problems & Schemes based on LWE~\cite{regev2009}, Module-LWE, or approximate lattice problems with formal reduction-based security arguments \\
\bottomrule
\end{tabular}
\end{table}

We note that this three-part classification deliberately avoids centering the framework on the complexity class NP. While NP-hardness of exact SVP is an important complexity-theoretic result, the security of deployed PQC schemes such as ML-KEM~\cite{nist_fips203} relies on \emph{approximate} lattice problem hardness and worst-case to average-case reductions~\cite{regev2009,peikert2016}, which is a distinct and more nuanced notion. NP appears in this framework only as part of the broader complexity-theoretic landscape, not as a claim that deployed PQC security is ``based on NP'' in any direct formal sense.

\subsection{Definitions}

We provide standard definitions for reference:

\begin{definition}[P --- Polynomial Time]
The class of decision problems solvable in polynomial time by a deterministic Turing machine.
\end{definition}

\begin{definition}[BQP --- Bounded-Error Quantum Polynomial Time]
The class of decision problems solvable in polynomial time by a quantum Turing machine with error probability bounded below a fixed constant.
\end{definition}

\begin{remark}[On BQP vs.\ NP]
The exact relationship between BQP and NP is unknown~\cite{aaronson2010}. There exist oracle separations in both directions. This uncertainty is precisely why the framework uses complexity classes as an \emph{interpretive language} rather than as a formal classification.
\end{remark}

\begin{remark}[On Lattice Problem Hardness]
The hardness assumptions underlying ML-KEM~\cite{nist_fips203} and ML-DSA~\cite{nist_fips204} concern approximate variants of lattice problems---specifically, Module-LWE---rather than exact SVP. The formal security foundations of deployed lattice-based schemes are rooted in reduction-based hardness arguments for LWE-type problems and related lattice assumptions; in deployed standards, these appear in module-lattice settings rather than the exact formulations of classical worst-case SVP.
\end{remark}

\subsection{Worked Example: RSA vs.\ ML-KEM}

To demonstrate the framework's operational use, we apply Definition~\ref{def:framework} to two concrete schemes. Table~\ref{tab:worked_example} shows the resulting interpretive security profiles.

\begin{table}[ht]
\centering
\caption{Worked example: Interpretive security profiles for RSA and ML-KEM under the Explainable PQC framework.}
\label{tab:worked_example}
\begin{tabular}{@{}lccc@{}}
\toprule
\textbf{Scheme} & \textbf{Classical ($S_c$)} & \textbf{Quantum ($S_q$)} & \textbf{Reduction-Backed ($S_r$)} \\
\midrule
RSA-2048 & Yes & No (Shor~\cite{shor1997}) & No \\
ML-KEM-768 & Yes & No known attack & Yes (Module-LWE~\cite{regev2009}) \\
\bottomrule
\end{tabular}
\end{table}

RSA-2048 is classically secure because no known polynomial-time factoring algorithm exists, but it is \emph{not} quantum secure because Shor's algorithm~\cite{shor1997} solves integer factorization in quantum polynomial time. Its security is also not reduction-backed in the worst-case to average-case sense: the hardness of factoring is an unstructured assumption without known worst-case reductions. ML-KEM-768~\cite{nist_fips203}, by contrast, faces no known classical or quantum polynomial-time attack, and its security \emph{is} reduction-backed: the Module-LWE problem on which it relies enjoys worst-case to average-case hardness guarantees rooted in lattice problem hardness~\cite{regev2009,peikert2016}.

\subsection{Limitations}

This classification does not replace formal security analysis. It does not provide concrete security estimates, does not analyze specific parameter choices, and does not account for side-channel attacks or implementation vulnerabilities. Its purpose is to provide an accessible entry point for understanding the \emph{type} of computational hardness that PQC relies upon.

\section{Exploratory Mathematical Investigation: Combinatorial Hodge Theory and Lattice Geometry}
\label{sec:math}

This section describes an exploratory mathematical investigation conducted by Rizwan Jahangir, aimed at developing structural understanding of lattice geometry from the perspective of combinatorial Hodge theory. This work was developed independently of the interpretive framework and may have implications for understanding lattice problems, though establishing a direct connection to computational complexity remains ongoing.

\subsection{Motivation: Structural Understanding of Lattice Hardness}

The security of lattice-based cryptography depends on the difficulty of finding short vectors in high-dimensional lattices. Shor's algorithm~\cite{shor1997} succeeds against RSA because integer factorization possesses algebraic periodicity exploitable by quantum Fourier transforms. A natural question is whether analogous exploitable structures exist in lattice problems. No polynomial-time quantum algorithms are currently known for the lattice problems underlying deployed PQC~\cite{peikert2016}.

Our investigation seeks to develop mathematical language and tools---grounded in combinatorial Hodge theory, toric varieties, polyhedral fans, and cohomological invariants---that may clarify the geometric landscape in which lattice hardness resides.

From a computer science perspective, the relevance of this investigation lies in the possibility of identifying structural features of lattice instances that correlate with algorithmic tractability or resistance. Even without current complexity-theoretic consequences, such structural decompositions may eventually help frame questions about whether certain lattice families exhibit differing algorithmic behavior, or contribute to refining explanatory models of why existing attacks (such as BKZ~\cite{schnorr1994}) fail beyond certain dimensions.

\subsection{Algorithmic Context: Structural Decomposition and Lattice Attacks}

To situate this mathematical investigation within the algorithmic landscape, we briefly review the main attack strategies against lattice problems. Existing attacks include BKZ-style basis reduction~\cite{schnorr1994,chen2011}, enumeration methods~\cite{gama2008}, and sieving algorithms. These approaches exploit geometric properties of lattice bases and Gram--Schmidt orthogonalization structures. The empirical behavior of these algorithms---in particular, the gap between theoretical worst-case bounds and observed practical performance~\cite{gama2008}---remains incompletely understood.

A structural decomposition of lattice fans, as studied in this section, may provide an alternative perspective for understanding why certain reduction strategies succeed locally (within small block sizes) but fail to propagate improvements globally (across the full lattice dimension). This connects the mathematical investigation to a concrete algorithmic question.

\paragraph{Quantum algorithms for lattice problems.} No polynomial-time quantum algorithms are currently known for the lattice problems underlying deployed PQC. The best known quantum approaches---including quantum variants of sieving and enumeration---provide at most quadratic improvements in time complexity over their classical counterparts, leaving the fundamental exponential scaling intact~\cite{peikert2016}. This contrasts sharply with Shor's algorithm for factoring, which achieves an exponential speedup. The absence of known efficient quantum algorithms for lattice problems is a central reason for the confidence in lattice-based PQC.

\subsection{Combinatorial Hodge Theory and Lattice Fans}

The work by Jahangir~\cite{jahangir2025} investigates the intersection cohomology of fans associated with toric varieties and its connection to the Hodge conjecture. In the context of this framework, the relevant question is whether the fan structure of lattices admits algebraic or combinatorial structures that could, in principle, inform our understanding of computational hardness. This is a question about \emph{structural understanding of lattice geometry}, not a claim of discovering quantum shortcuts.

\subsection{Local Generation Theorem: Decomposition of Lattice Relations}

Jahangir~\cite{jahangir2026} presents the local generation theorem, which decomposes the problem of analyzing relations in a lattice fan into local subproblems. The theorem proves that the relations of a lattice fan can be decomposed into local neighborhood structures called ``stars,'' and the global structure can be reconstructed from these local pieces.

This decomposition provides a tool for distinguishing whether the difficulty of lattice problems originates from the complexity of local structures or from global topological connectivity---a distinction that could be valuable for understanding the hardness landscape.

\begin{hypothesis}[Local Structure]
\label{hyp:local}
If the relations of a lattice fan can be decomposed into bounded-size local structures (stars), then algorithmic attacks relying on local reduction steps---such as block-wise BKZ reduction~\cite{schnorr1994,chen2011}---may fail to capture global lattice geometry. This suggests that global connectivity, rather than local complexity alone, may contribute to the observed hardness of lattice problems at high dimensions.
\end{hypothesis}

This hypothesis does not constitute a complexity-theoretic proof; it is a \emph{conjectural structural statement} that connects the mathematical decomposition to an algorithmic question. Testing it---for instance, by examining whether lattice instances with differing global connectivity exhibit measurably different BKZ performance---is a direction for future empirical and theoretical work.

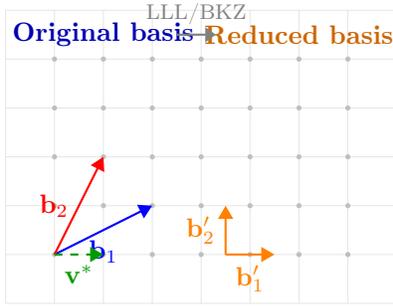
\begin{figure}[ht]
\centering
\begin{tikzpicture}[scale=0.65]
\draw[gray!20, very thin] (-1,-1) grid (7,5);

\foreach \x in {0,1,...,6} {
    \foreach \y in {0,1,...,4} {
        \fill[gray!50] (\x,\y) circle (1.5pt);
    }
}

\draw[-{Triangle[length=5pt]}, thick, blue] (0,0) -- (2,1) node[midway, below, font=\small] {$\mathbf{b}_1$};
\draw[-{Triangle[length=5pt]}, thick, red] (0,0) -- (1,2) node[midway, left, font=\small] {$\mathbf{b}_2$};

\draw[-{Triangle[length=5pt]}, thick, green!60!black, dashed] (0,0) -- (1,0) node[midway, below, font=\small] {$\mathbf{v}^*$};

\draw[-{Triangle[length=5pt]}, thick, orange] (3.5,0) -- (4.5,0) node[midway, below, font=\small] {$\mathbf{b}_1'$};
\draw[-{Triangle[length=5pt]}, thick, orange] (3.5,0) -- (3.5,1) node[midway, left, font=\small] {$\mathbf{b}_2'$};

\node[font=\small\bfseries, blue!70!black] at (1, 4.5) {Original basis};
\node[font=\small\bfseries, orange!80!black] at (5, 4.5) {Reduced basis};
\draw[-{Triangle[length=4pt]}, thick, gray] (2.5, 4.5) -- (3.3, 4.5) node[midway, above, font=\scriptsize] {LLL/BKZ};

\end{tikzpicture}
\caption{Schematic illustration of lattice basis reduction. A lattice (gray points) with an original basis $(\mathbf{b}_1, \mathbf{b}_2)$ is transformed via reduction algorithms (LLL/BKZ) into a reduced basis $(\mathbf{b}_1', \mathbf{b}_2')$ with shorter, more orthogonal vectors. The dashed vector $\mathbf{v}^*$ represents the shortest lattice vector (SVP solution). In high dimensions, finding $\mathbf{v}^*$ becomes computationally intractable.}
\label{fig:lattice_geometry}
\end{figure}

\subsection{Summary of Status}

To clearly delineate the status of this investigation:
\begin{itemize}[leftmargin=*]
    \item \textbf{Established mathematical result:} The local generation theorem for rational fans~\cite{jahangir2026}, which provides a decomposition of lattice fan relations into local data.
    \item \textbf{Interpretive claim:} Such decompositions may help reason about the geometric structure underlying lattice hardness, potentially informing future algorithmic or complexity-theoretic analysis.
    \item \textbf{Open problem:} Whether these geometric and cohomological structures imply anything about the computational complexity of lattice problems or the security of lattice-based cryptography.
\end{itemize}
No computational shortcut for lattice problems has been discovered or claimed. Several additional bridging results would be required to draw cryptographic conclusions.

\section{Empirical Layer: Open-Source Lattice Solver Verification Platform in Julia}
\label{sec:implementation}

As the empirical layer supporting the engineering validity of this research, we constructed an open-source framework called ``Open Problem Toolkit''\footnote{\url{https://github.com/Kiara-01-Lab/open-problem-toolkit}} using the Julia programming language. The purpose of this toolkit is to numerically quantify the theoretical ``intractability (hardness)'' of lattice-based cryptography in terms of actual computational resources (processing time, memory usage, computational complexity), and to provide objective and reproducible evidence. We intentionally limited the scope to computationally tractable low-dimensional lattices, constructing both lattice cryptography builders and solvers within this range.

\subsection{Background and Objectives of Development}

This toolkit performs a consistent pipeline from lattice construction, through basis reduction (a technique for reorganizing the structure of a complex lattice to search for decryption clues), to benchmark measurement. This makes it possible to reveal---through experimentation in low dimensions---the gap between ``theoretical verification'' and ``computational performance.''

\subsection{Package Architecture}

The framework consists of the following four modules, and verification can be executed independently for each.

\begin{table}[ht]
\centering
\caption{Package architecture of the Open Problem Toolkit.}
\label{tab:packages}
\begin{tabular}{@{}lp{3.5cm}p{6.5cm}@{}}
\toprule
\textbf{Package Name} & \textbf{Functional Overview} & \textbf{Role Description} \\
\midrule
\texttt{LatticeReductionAlgorithms.jl} & Verification of lattice decryption capability using LLL and KZ methods & Simplifies the lattice structure to make it easier to find the shortest vector (the cryptographic key). \\
\addlinespace
\texttt{LatticeBasedCryptography.jl} & Implementation of various lattice-based cryptographic schemes & Performs encryption and decryption operations to verify correct functionality. \\
\addlinespace
\texttt{PQCValidator.jl} & Verification of security boundaries & Determines the dimension threshold (lattice complexity) required to be considered secure. \\
\addlinespace
\texttt{LibOQS.jl} & Integration with the NIST standard library & Calls world-standard algorithms (in C) from Julia for comparative verification. \\
\bottomrule
\end{tabular}
\end{table}

\subsection{Experimental Results and Discussion}

Using the constructed verification platform, we measured the performance of algorithms that solve the Shortest Vector Problem (SVP), which lies at the core of lattice-based cryptography.

\subsection{Implementation Environment: Why the Julia Language?}

In this study, we adopted Julia as the experimental language. Evaluating the security of lattice-based cryptography requires enormous iterative computations; Julia features the ease of writing characteristic of Python while achieving execution speeds comparable to C.

\subsection{``Computational Explosion'' with Increasing Dimension and Changes in Solver Feasibility}

Using a MacBook equipped with an Apple silicon chip, we verified the success or failure of lattice cryptography solvers starting from low dimensions.

\begin{enumerate}[leftmargin=*]
    \item \textbf{Lattice Cryptography Solver Verification at 10 Dimensions:}
    \begin{itemize}
        \item Result: Both LLL (approximate solver)~\cite{lenstra1982} and EKZ (exact solver) successfully completed their computations.
        \item Discussion: Lattice cryptography at this scale can be easily broken by a standard CPU-based PC and is not secure.
    \end{itemize}

    \item \textbf{Lattice Cryptography Solver Verification at 40 Dimensions:}
    \begin{itemize}
        \item Result: The LLL approximate solver succeeded, but the EKZ exact solver timed out and failed to complete.
        \item Discussion: Even at 40 dimensions---which is very small compared to NIST cryptographic standards (400--500+ dimensions)---the cost of finding an exact solution grows at a certain exponential rate, reaching CPU computational limits even with Julia.
    \end{itemize}
\end{enumerate}

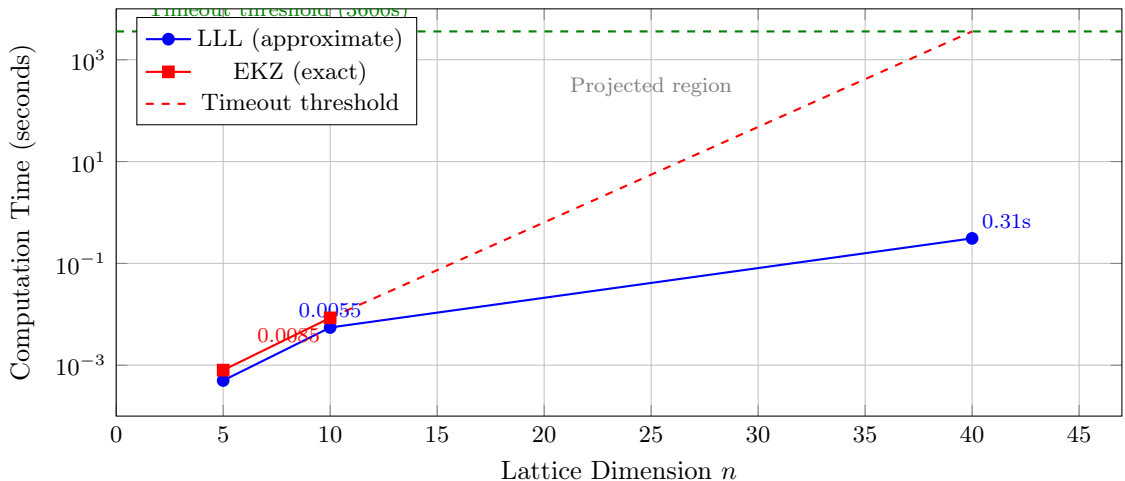
\begin{figure}[ht]
\centering
\begin{tikzpicture}
\begin{semilogyaxis}[
    width=0.85\textwidth,
    height=7cm,
    xlabel={Lattice Dimension $n$},
    ylabel={Computation Time (seconds)},
    xmin=0, xmax=47,
    ymin=1e-4, ymax=1e4,
    legend style={font=\footnotesize, at={(0.02,0.98)}, anchor=north west},
    grid=both,
    grid style={line width=0.1pt, draw=gray!30},
    major grid style={line width=0.2pt,draw=gray!50},
    tick label style={font=\footnotesize},
    label style={font=\small},
]
\addplot[blue, mark=*, thick] coordinates {
    (5, 0.0005)
    (10, 0.0055)
    (40, 0.31)
};
\addlegendentry{LLL (approximate)}

\addplot[red, mark=square*, thick] coordinates {
    (5, 0.0008)
    (10, 0.0085)
};
\addlegendentry{EKZ (exact)}

\addplot[red, dashed, thick] coordinates {
    (10, 0.0085)
    (40, 3600)
};

\addplot[green!50!black, dashed, thick, domain=0:47] {3600};
\addlegendentry{Timeout threshold}

\node[font=\scriptsize, gray] at (axis cs:25,300) {Projected region};
\node[font=\scriptsize, blue, anchor=south] at (axis cs:10,0.0055) {0.0055};
\node[font=\scriptsize, blue, anchor=south west] at (axis cs:40,0.31) {0.31s};
\node[font=\scriptsize, red, anchor=north east] at (axis cs:10,0.0085) {0.0085};
\node[font=\scriptsize, green!50!black, anchor=south west] at (axis cs:1,3600) {Timeout threshold (3600s)};

\end{semilogyaxis}
\end{tikzpicture}
\caption{Computation time versus lattice dimension $n$ for LLL (approximate) and EKZ (exact) solvers. The EKZ solver times out at 40 dimensions, illustrating the rapid growth of exact-solution computation cost even at low dimensions.}
\label{fig:computation_time}
\end{figure}

\subsection{Conclusion: Correlation Between Theory and Measurement}

Through these experiments, we partially demonstrated the security foundation of lattice-based cryptography---the ``difficulty of exact-solution computation in high dimensions.'' Even when an attacker uses the efficient LLL approximate solver, there is a gap between the approximate solution and the actual exact solution needed for a true solver; and the EKZ exact solver, even on a CPU, hits a wall at low dimensions. This makes it possible to design experimental paths toward verification using actual quantum computers. Furthermore, by using the Open Problem Toolkit, students and industry members can conduct hands-on experiments and gain understanding of low-dimensional lattice cryptography (which we call ``educational TOY LATTICE'') on personal PCs, potentially advancing concrete understanding of quantum security in society.

\section{Threats to Validity and Scope Limitations}
\label{sec:limitations}

We explicitly acknowledge the following limitations of this work:

\begin{itemize}[leftmargin=*]
    \item \textbf{The framework is not a proof system.} The three-part interpretive layer (Section~\ref{sec:theory}) does not provide formal security guarantees. It is a communication and reasoning tool, not a replacement for reduction-based cryptographic analysis~\cite{regev2009,peikert2016}.
    \item \textbf{No concrete security estimates are provided.} The framework does not analyze specific parameter choices, side-channel resistance, or implementation-level security of deployed standards such as ML-KEM~\cite{nist_fips203}.
    \item \textbf{The mathematical layer is exploratory.} The combinatorial Hodge theory investigation (Section~\ref{sec:math}) is not yet connected to computational complexity theory. No cryptographic implications have been established.
    \item \textbf{The empirical layer is illustrative.} The experiments in Section~\ref{sec:implementation} are limited to low-dimensional lattices (10D--40D) that are far below the parameter ranges of deployed cryptographic standards (typically 400--500+ dimensions). These experiments are pedagogical and illustrative of scaling behavior, not security evaluations of real systems.
    \item \textbf{Audience scope.} The framework is designed for researchers and industry practitioners seeking interpretability of PQC assumptions. It is not intended for use in formal security certification or compliance processes.
\end{itemize}

\section{Conclusion and Future Work}
\label{sec:conclusion}

\subsection{Summary of Contributions}

This paper proposed ``Explainable PQC,'' a layered interpretive framework for communicating post-quantum cryptographic security assumptions. The framework comprises:
\begin{enumerate}[leftmargin=*]
    \item A \textbf{three-part security interpretation} (Section~\ref{sec:theory}) providing a structured vocabulary for distinguishing classical vulnerability, quantum vulnerability, and reduction-backed post-quantum assumptions---with an explicit formal statement of scope (Proposition~\ref{prop:scope}).
    \item An \textbf{exploratory mathematical investigation} (Section~\ref{sec:math}) applying combinatorial Hodge theory and the local generation theorem to study lattice geometry, with a clear separation of established mathematical results, interpretive claims, and open problems.
    \item An \textbf{empirical experimentation platform} (Section~\ref{sec:implementation}) illustrating the rapid scaling of exact lattice-solver computation cost in the tested low-dimensional regime.
\end{enumerate}

\subsection{Future Research Directions}

Future work will pursue three directions:
\begin{enumerate}[leftmargin=*]
    \item \textbf{Mathematical bridging:} Developing the theoretical results needed to connect the geometric and cohomological structures studied in Section~\ref{sec:math} to the algorithmic complexity of lattice problems, including potential connections to known lattice reduction algorithms such as BKZ~\cite{schnorr1994}.
    \item \textbf{Higher-dimensional experimentation:} Extending empirical measurements to higher-dimensional lattices and to more capable compute environments, with the long-term goal of studying settings closer to deployed standards~\cite{nist_fips203}.
    \item \textbf{Framework refinement:} Incorporating concrete security estimation methodologies and recent results on quantum algorithms for lattice problems into the interpretive model.
\end{enumerate}

\paragraph{Potential evaluation methodology.} Future empirical validation of this framework could involve systematic benchmarking of lattice reduction algorithms across increasing lattice dimensions and structured lattice families (e.g., ideal lattices vs.\ random lattices). Such experiments could examine whether geometric features identified through combinatorial decompositions correlate with solver performance, providing a falsifiable test of the structural hypotheses advanced in Section~\ref{sec:math}.

The central claim of this paper is therefore modest but precise: while formal PQC security continues to depend on reduction-based cryptographic analysis, there is value in developing structured interpretive frameworks that relate complexity-theoretic models, mathematical structure, and implementation evidence. We view Explainable PQC as a step toward such a framework.

\bibliographystyle{unsrt}

\end{document}